\documentclass[twocolumn,showpacs,preprintnumbers,amsmath,amssymb,prb,superscriptaddress]{revtex4-1}
\usepackage{amsfonts}
\usepackage[english]{babel}
\usepackage[T1]{fontenc}
\usepackage[usenames]{color}
\usepackage{times}
\usepackage{mathrsfs}
\usepackage{graphicx}
\usepackage{dcolumn}
\usepackage{bm}
\usepackage[colorlinks,bookmarks=true,citecolor=blue,linkcolor=red,urlcolor=blue]{hyperref}
\usepackage[tight, FIGTOPCAP, hang, raggedright, nooneline]{subfigure}
\usepackage{hyperref}
\usepackage{mathbbol}
\usepackage{bbm}

\newtheorem{theorem}{Theorem}
\newtheorem{lemma}[theorem]{Lemma}

\newtheorem{observation}[theorem]{Observation}

\newcommand{\je}[1]{\textcolor{black}{#1}}
\newcommand{\ejb}[1]{\textcolor{black}{#1}}

\newcommand\RR{{\mathbbm{R}}}

\newcommand\id{{\mathbbm{1}}}

\newcommand{\e}{{{\rm e}}}
\newcommand{\dist}{{{\rm d}}}

\newcommand{\proofend}{\hfill\fbox\\\medskip }

\DeclareMathOperator{\tr}{tr}

\begin{document}

\title{Topological insulators with arbitrarily tunable entanglement}
\author{J.\ C.\ Budich}
\affiliation{Department of Physics, Stockholm University, SE-106 91 Stockholm, Sweden}

\affiliation{Institute for Theoretical Physics, University of Innsbruck, 6020 Innsbruck, Austria}
\affiliation{Institute for Quantum Optics and Quantum Information, Austrian Academy of Sciences, 6020 Innsbruck, Austria}
\author{J.\ Eisert}
\affiliation{Dahlem Center for Complex Quantum Systems,
Freie Universit\"at Berlin,
Arnimallee 14, 14195 Berlin, Germany}
\author{E.\ J.\ Bergholtz}
\affiliation{Dahlem Center for Complex Quantum Systems,
Freie Universit\"at Berlin,
Arnimallee 14, 14195 Berlin, Germany}
\date{\today}
\begin{abstract}
We elucidate how Chern and topological insulators fulfill an area law for the entanglement entropy. By explicit construction of a family of lattice Hamiltonians, we are able to demonstrate that the area law contribution can be tuned to an arbitrarily small value, but is topologically protected from vanishing exactly. We prove this by introducing novel methods to bound entanglement entropies from correlations using perturbation bounds, drawing intuition from ideas of quantum information theory. This rigorous approach is complemented by an intuitive understanding in terms of entanglement edge states. These insights have a number of important consequences: {The area law has no universal component, no matter how small, and the 
entanglement scaling cannot be used as a faithful diagnostic of topological insulators.}
{This holds for all Renyi entropies which uniquely determine the entanglement spectrum \ejb{which is hence also non-universal}.} The existence of arbitrarily weakly entangled topological insulators {furthermore} opens up possibilities of devising correlated topological phases in which the entanglement entropy is small and which are thereby numerically tractable, specifically in tensor network approaches.
\end{abstract}
\maketitle
\section{Introduction and key results}
Since the experimental discovery \cite{Klitzing1980,StormerFQH} and first theoretical explanation \cite{Laughlin1981,TKNN1982,LaughlinState} of the quantum Hall effect, topological phenomena have triggered some of the most active {and} intriguing research fields in physics. An important milestone in the theoretical understanding of topological phases has been reached with the notion of topological order \cite{WenTO} as a means to distinguish phases of matter beyond the the paradigm of local order parameters associated with spontaneous symmetry breaking \cite{AndersonBasic}.
In parallel, the entanglement entropy which measures the amount of quantum correlations between a system and its environment as a function of the boundary `area' $L$ has developed to a standard tool in quantum many body physics, providing important insights into properties of a wide range of physical systems \cite{JensReview}. Notably, it can be used to distinguish, sometimes even classify, different phases of matter thus providing insights that are valuable, e.g. in the context of numerical simulations \cite{DMRGMPS}.

More recently, the notions of entanglement scaling and topological phases have been bridged showing that there is a topological contribution $\gamma$  to the entanglement entropy of a bipartite system that is unique to topologically ordered phases.
This scale invariant term has been coined {\it topological entanglement entropy} (TEE) and depends only on the logarithm of the total quantum dimension of the topological phase
\cite{KitaevPreskill, LevinWenEntropy,Zanardi,Zanardi2}. 
Putting together these concepts, the generic scaling of the entanglement entropy in two spatial dimensions (2D) reads as
\begin{align}\label{scaling}
S{(L)}= \alpha L-\gamma + O (1/L){,} 
\end{align}
{for a suitable {$\alpha \geq 0$}.}
However, the whole family of integer quantum Hall states \cite{Klitzing1980,Laughlin1981,TKNN1982} has quantum dimension one implying a vanishing TEE $\gamma$. Still, integer quantum Hall states and their lattice translation invariant analogues called {\it Chern insulators} (CIs) 
\cite{QAH} are gapped topological phases that are not characterized by any conventional local order and cannot be adiabatically connected to trivial insulators. In fact, according to the definition suggested in Ref.\ \cite{WenLU}, the CIs belong to the class of topologically ordered systems.

\begin{figure}[th]
\includegraphics[width=0.78\linewidth]{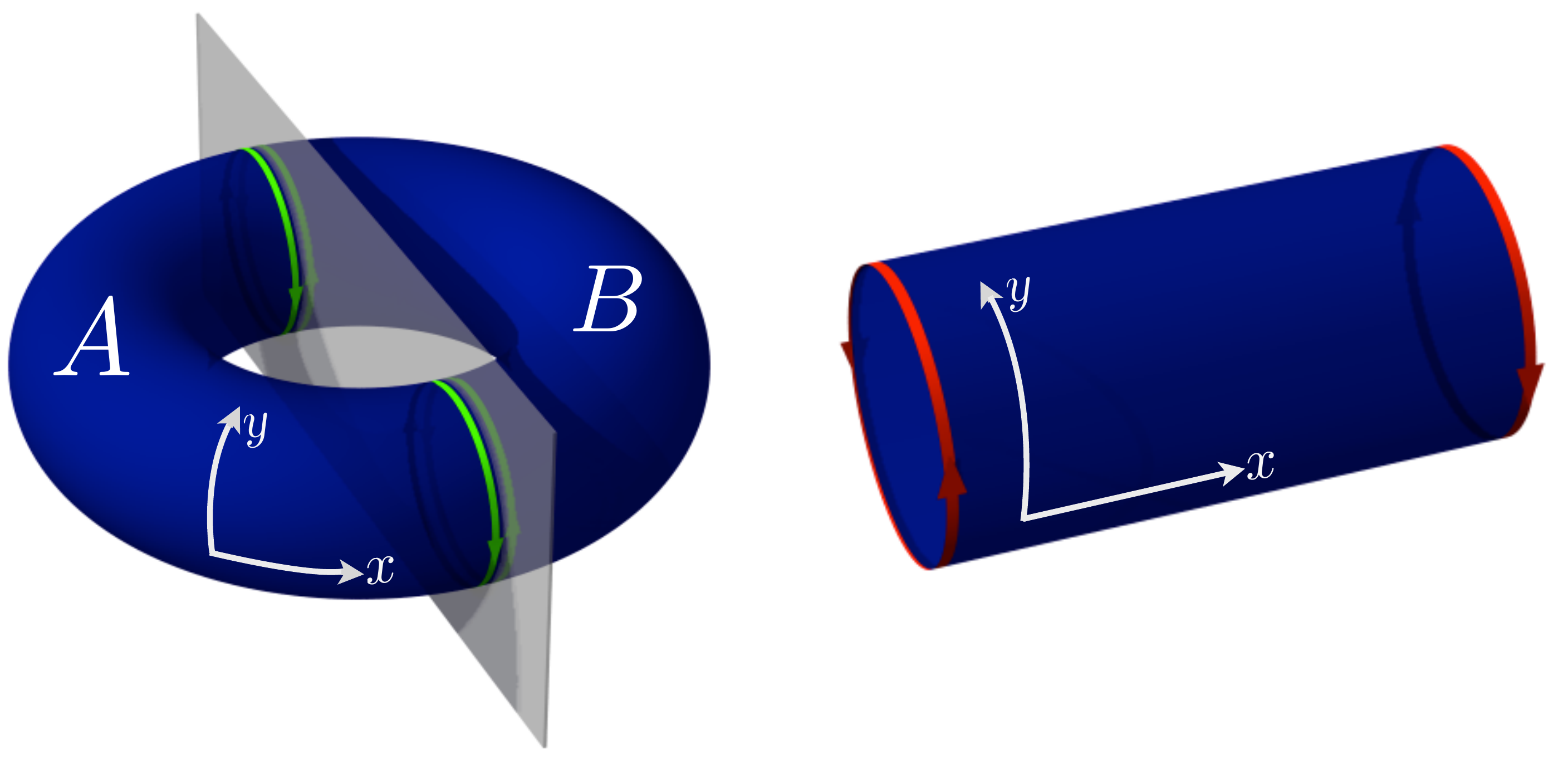}
\caption{(Color online) Sketch of our setup and the bulk-boundary correspondence in Chern insulators. Bi-partitioning the system into distinct regions ($A$ and $B$) on a torus gives rise to virtual edge states (green) in the entanglement Hamiltonian which are close analogues of the physical edge states (red) on a cylinder.}
\label{fig:cylindercut}
\end{figure}

In this work, we pose the natural and important question to what extent the topological nature of CI states and topological insulators in 2D \cite{KaneMele2005a,KaneMele2005b,BHZ2006} can be inferred from their entanglement scaling, {and more generally, entanglement spectra}. 
Our analysis, {expected to} generalize to all standard topological insulator classes \cite{Schnyder2008,KitaevPeriodic,RyuLudwig} with spatial dimension $d>1$, is rooted in the following observations. {The} area law stems from two qualitatively different contributions: A trivial contribution and a topological one. The topological part is in one to one correspondence to the topologically protected edge states occurring at the boundary of a CI (Fig.\ \ref{fig:cylindercut}). This contribution {may} be considered a fingerprint of the non-vanishing Chern number in the sense that its value will be non-zero for any state that is adiabatically connected to a non-trivial CI. {Remarkably, however, we find that there is no non-zero 
lower bound, which might have been expected in  analogy with
the Lowest Landau level, which is a Chern insulator enjoying minimum fluctuations in the (guiding center) coordinates which fail to commute, $[X,Y]=i\ell_B^2$, when projected to the band.  In contrast, the trivial part can be adiabatically tuned to zero just as the entire entanglement can be tuned to zero for trivial insulators in the atomic limit. In order to alter both {contributions} 
simultaneously, we devise a novel construction of a Chern insulator model built via `dimensional extension' \cite{WZW,QiTFT} of a topologically non-trivial one-dimensional (1D) model.  We are able to {demonstrate} that no {lower bound can be put on $\alpha$ even for a CI state and that it has no even arbitrarily small
universal component}. {The local correlations can be chosen arbitrarily close to those of uncorrelated Slater determinants.} In fact, this is true for the whole family of so-called Renyi entropies, $S_p(L)$, each having a scaling of the form of Eq. (\ref{scaling}). The collection of (integer) Renyi entropies fully determine the entanglement spectrum \cite{LiHaldane}, and our results imply that even this more complete information does not contain truly universal information beyond the fingerprint mentioned above.

{In order to prove this we introduce rigorous methods of upper bounding entanglement entropies {in terms of} {\it correlations} only, using perturbation bounds on spectra of fermionic {correlation} 
matrices {and instruments from harmonic analysis}. {On an intuitive level,} we show that the contribution to $\alpha$ depends 
on the steepness of the dispersion of the edge states in a sample with fixed boundaries. {This intuition is guided by the insight that the bulk entanglement spectrum \cite{LiHaldane} is typically 
closely related to the gapless excitations at the edge of a topological state \cite{LiHaldane,Qi}.} 
The possibility of arbitrarily suppressing the area law coefficient may be useful for the construction of representatives of correlated topological phases like fractional Chern insulators \cite{kapitmueller,wenfci,sunfci,neupertfci,EmilReview},
specifically with tensor network methods where the pre-factor of the area law 
relates to the required bond dimension.

The remainder of this article is organised as follows: in Section \ref{sec:freeferm}, we generally discuss entanglement properties of free fermonic systems with a focus on the relation between topological edge states and the area law coefficient of the entanglement scaling. Building on this general analysis, a family of Chern insulators with arbitrarily tunable entanglement is constructed in Section \ref{sec:tunableCI}. Upper and lower bounds on the Renyi entanglement entropies of this model family are obtained from rigorous analytical methods introduced in this work (for details see the appendix) and confirmed by extensive numerical analysis. Finally, we present a concluding discussion in Section \ref{sec:conclusion}.

\section{Entanglement in free fermonic systems}
\label{sec:freeferm}
\subsection{Two-banded fermionic systems and entanglement Hamiltonians}
The CI models we {are} considering are non-interacting gapped {two-banded} fermionic systems {with no pairing terms}. {For two-dimensional cubic lattices with $L\times L$ sites on a torus, such a Hamiltonian takes the form 
\begin{equation}
	H = \sum_{I,J} c_I^\dagger  h_{I,J} c_J ,
\end{equation}
where the fermionic modes are labeled by $I=(j,k,\updownarrow)$ for $j,k=1,\dots, L$.}
Their ground states are Slater determinants of all single particle states below the energy gap. When the system is divided into two subsystems $A$ and $B$ by virtue of a cut in real space 
(Fig.\ \ref{fig:cylindercut}), the reduced state of the individual subsystems generically exhibits a non-zero entanglement entropy. 
{Ground states $\rho$ of such models are always fermionic Gaussian states. This 
implies that} the reduced density matrix $\rho_A$ for subsystem $A$ can be viewed as a free fermionic thermal state with {unit inverse temperature} 
of the isolated subsystem $A$, i.e.,
\begin{equation}
\rho_A= {\e^{-H_E}}/{\tr (\e^{-H_E})}
\end{equation}
{where the {\it entanglement Hamiltonian} 
$H_E$ is again quadratic in the fermionic operators.}
For such free fermionic models the ground state $\rho $ is defined by the  Hermitian, {positive} 
correlation matrix $C$, with entries
\begin{equation}
	C_{I,J} = \tr(\rho c_I^\dagger c_J).
\end{equation}
$H_E$ is determined in terms of the truncated correlation matrix, $C^{(A)}$, {the sub-matrix of $C$ associated with indices only in $A$.}
If $\left\{\xi_j\right\}$ denotes the set of eigenvalues of $C^{(A)}$ and $\left\{\epsilon_j\right\}$ the set of single particle entanglement energies, the
relation reads for all $j$ as
\cite{Peschel2003}
\begin{align}
\epsilon_j= \log{{(1-\xi_j)}{\xi_j^{-1}}}.
\label{eqn:chrelation}
\end{align}

\subsection{Entanglement entropies.} 
{
In momentum space,
 the Hamiltonians become
\begin{equation}\label{model}
	H = \sum_{\mathbf k} (f_{\uparrow}^\dagger({\mathbf k}),f_{\downarrow}^\dagger({\mathbf k})) h({\mathbf k})
	\left(
	\begin{array}{c}
	f_{\uparrow}({\mathbf k})\\
	f_{\downarrow}({\mathbf k})
	\end{array}
	\right)
\end{equation}
where ${\mathbf k}=(k_x,k_y)$,
\begin{equation}
	h({\mathbf k}) = \sum_{j=1}^3 d_j({\mathbf k}) \sigma_j,
 \end{equation}
and $\sigma_1,\sigma_2,\sigma_3$ are the Pauli matrices. The wave vectors take the values $k_x,k_y=2\pi l/L\in(-\pi,\pi]$ for $l=-L/2+1,\dots, L/2$
({we also allow for $L_x\times L_y$-lattices}).
The two energy bands are separated by $ \pm\|{\mathbf d}({\mathbf k})\|$.  
For a fully occupied lower and unoccupied upper band, the correlation matrix in momentum space of the ground state 
is 
\begin{equation}
	\bar{C} = \bigoplus_{\mathbf k} \frac{1}{2}\biggl(\id - \sum_{j=1}^3 \frac{d_j({\mathbf k})}{\| {\mathbf d}\| } \sigma_j\biggr).
\end{equation}
Given the toroidal symmetry of the problem, when computing the entanglement entropy, we keep the momentum space along the $k_y$ direction,
but turn to real space otherwise \cite{JensStat}. The fermionic operators are then transformed as
\begin{equation}
{f_{\updownarrow}({\mathbf k})=  \sum_{l=1}^{L} \e^{-i k_x l} c_{l, \updownarrow}(k_y)/L^{1/2}}.
\end{equation}
The ground state of each decoupled Hamiltonian labeled by $k_y$ is associated with a correlation matrix $C({k_y})$ in real space. 
\je{We allow for arbitrary Renyi entanglement entropies $S_p$, $p\geq 1$, the standard von-Neumann entropy being recovered in the limit
$p\downarrow 1$. Indeed, it is known that all integer Renyi entropies uniquely determine the entire entanglement spectrum. Hence, our study allows
to conclude that the entanglement spectrum is also non-universal.}
The entanglement entropy $S_p$
decouples into a sum over the contributions for each $k_y$.
In this bi-partition, only sub-matrices of the correlation matrix 
contribute the indices of which belong to 
$A$. Denote with $C^{(A)}(k_y)$ the sub-matrix of $C(k_y)$ corresponding to sites contained in $A$, with eigenvalues 
$\{\xi_j (C^{(A)}(k_y))\}$. Then, defining 
\begin{equation}
	h_p(x)= {\log_2 (x^p+(1-x)^p )}/({1-p}),
\end{equation}
the expression for the Renyi entanglement entropy takes the form
\begin{equation}\label{EntropyFormula}
	\je{S_p{(L)} = \sum_{k_y}\sum_j h_p( \xi_{j} (C^{(A)}(k_y) ) }.
\end{equation}

\subsection{Analogy with edge states of a cylinder}
To get a physical intuition for the {situation at hand} it is helpful to consider  $Q=\id/2-C$.
It can be interpreted as a Hamiltonian with the same eigenstates as the original system but with flat bands, i.e., $\epsilon_-=-1/2$ for all occupied states and 
$\epsilon_+=1/2$ for all empty states \cite{Schnyder2008,Turner2009,Fidkowski,Turner2012,Hughes,FlatBandFootnote}. 
 Eq.\ (\ref{eqn:chrelation}) implies that the truncated flat band Hamiltonian $Q^{(A)}=\id/2-C^{(A)}$ is  related to the entanglement Hamiltonian as 
\begin{equation}
	H_E=2~\text{arctanh}(2Q^{(A)}) 
\end{equation}
\cite{Turner2009,Alexandradinata}.
It is thus clear that $H_E$ and the physical Hamiltonian $H$ must have similar properties regarding 
topologically protected edge states: $Q$ results from $H$ via adiabatic deformation and is hence topologically equivalent to the physical Hamiltonian.
$H_E$ is related to $Q^{(A)}$ via a monotonous mapping of its spectrum. 

The area law character of edge mode contributions to the entanglement entropy can be {intuitively} 
understood  considering a cylinder geometry where the cut is translation-invariant in the $y$-direction. A chiral edge state of $H_E$ is then described by 
{an energy dispersion}
$\epsilon_e$ of the momentum $k_y$ along the cut crossing the energy gap. The number of low lying entanglement levels associated with that edge state grows {linearly} with the length of the cut
$L$. 
The quantized $L$ wave vectors $k_y$ are equidistant so that the number of states in the edge mode dispersion satisfying $\epsilon_e(k_y)<\epsilon_c$ grows linearly with $L$ for an arbitrary cutoff $\epsilon_c>0$. Hence, a chiral edge mode results in a non-vanishing area law for the entanglement entropy, i.e., $S_p\geq \alpha L$ {for some $\alpha>0$}. 
The expected $\alpha$ contributed by the chiral edge mode can be {made plausible} at this simple level {(for a rigorous treatment, see below)}: If the edge state dispersion is steep, $\epsilon_e$ will cross the gap rapidly as a function of $k_y$ resulting in only a small fraction of its $L$  levels having low energies. A steep edge dispersion hence implies little entanglement.

\section{Chern insulators with tunable area law}
\label{sec:tunableCI}
\subsection{Model building}
We now construct a family of CI states with Chern number one in which the steepness of the edge states and with that the coefficient $\alpha$ of the area law can be tuned by a single parameter $\mu$. To this end we proceed in three steps. First, we discuss the entanglement scaling of a well known Dirac model for a CI \cite{QiTFT,RyuLudwig} from a viewpoint of dimensional extension. Second, we introduce a means to tune the topological edge state contribution to the area law to an arbitrary value. Third, we show how to get rid of the non-topological contribution to the {entanglement}
which otherwise masks the edge state contribution. Our analysis is inspired by the fact that 2D CI states can be obtained by dimensional extension \cite{WZW,QiTFT} of particle hole symmetry (PHS) preserving topologically non-trivial 1D band structures \cite{SSH,SSHReview,QiTFT}. In our case, $k_y$, the momentum variable along the cut, 
plays the role of the {additional} coordinate of the dimensional extension. At $k_y=0$ we define a 1D model as
\begin{align}
	h(k_x,k_y=0)=\sin(k_x)\sigma_{{2}}-\cos(k_x)\sigma_{{3}}
\label{eqn:ssh}
\end{align}
in Eq.\ (\ref{model}).
This 1D model is topologically characterized by a quantized Zak-Berry phase \cite{ZakPol} of $\pi$ which is protected by the PHS $\mathcal C= \sigma_{{1}} K$ ~\cite{HatsugaiQuantizedBerry}, where $K$ denotes complex conjugation. An interpolation with Chern number $1$ between Eq.\ (\ref{eqn:ssh}) and the trivial 1D model $h(kx,ky=\pm \pi)=\sin(k_x)\sigma_{{2}} +(2-\cos(k_x))\sigma_{{3}}$ 
is given by the Dirac model for a CI \cite{QiTFT}, i.e.,
\begin{eqnarray}\label{eqn:extensionSimple}
h(k_x,k_y)=&-&\sin(k_y) \sigma_{{1}} + \sin(k_x)\sigma_{{2}} \nonumber\\
&+&(1-\cos(k_y)-\cos(k_x))\sigma_{{3}}.
\end{eqnarray}
The nature of the edge states may be understood from the dimensional extension: The 1D model at $k_y=0$ supports a single pair of zero-energy end states, while the trivial model at $k_y=\pm \pi$ does not have any subgap states. During the gapped interpolation, these zero modes must hence be gapped out continuously which gives rise to a single chiral edge mode crossing the gap of the 2D model with fixed boundary conditions.

\begin{figure*}[ht]
\includegraphics[width=0.9\linewidth]{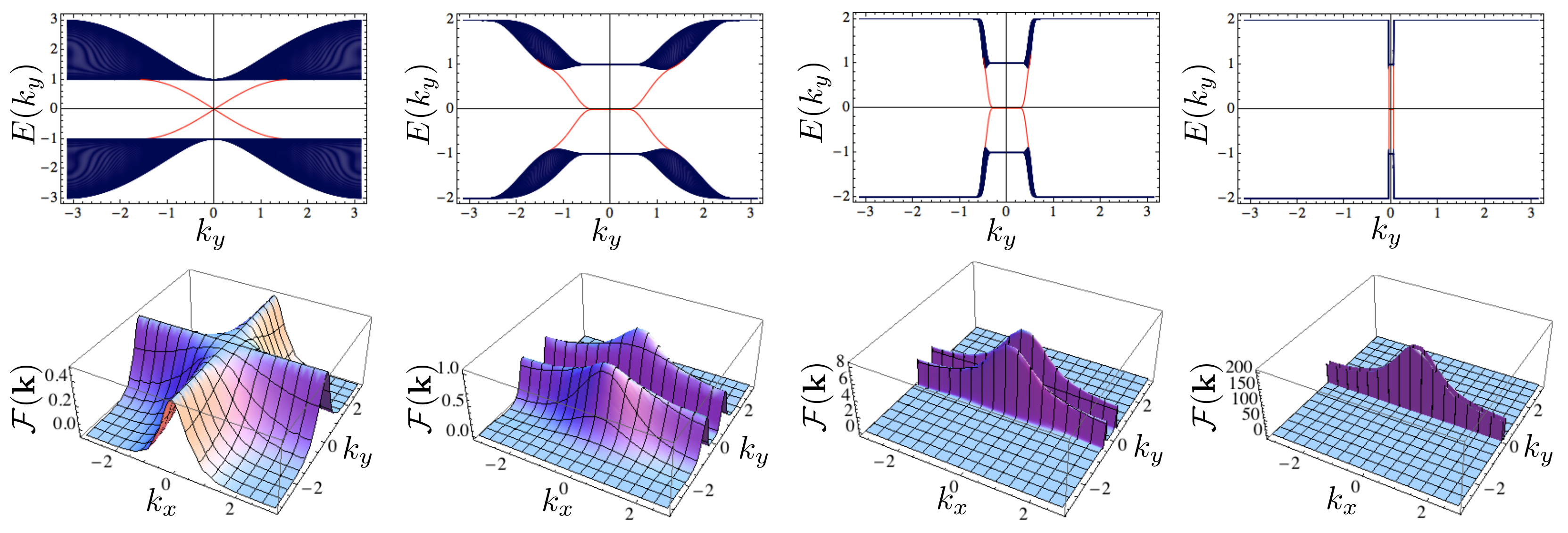}
\caption{(Color online) Band structure and Berry curvature. The upper panel shows the energy dispersion as a function of the transverse momentum $k_y$ {in the model defined on a cylinder
serving as an analogy}, and the lower panel shows the corresponding Berry curvature distribution \ejb{ $\mathcal{F}$} {for the model on the torus}. 
From the left to right, the data is displayed for the standard Dirac model (\ref{eqn:extensionSimple}) and our model (\ref{eqn:model}), with $\mu=\pi,1,0.2$, respectively.
{For small values of $\mu>0$, the Berry curvature is strongly peaked around $k_y=\pm \mu/2$ \ejb{which is necessary to maintain a unit Chern number,  $C= \int_{{\rm BZ}}\mathcal{F}(\mathbf k){\rm d}^2k/(2\pi)$}. The edge states exhibit a very steep slope around $k_y=\pm \mu/2$ and a plateau of width $\mu$ at zero energy between $k_y=-\mu/2$ and $k_y=\mu/2$.}}
\label{fig:edgeberry}
\end{figure*}

{To arrive at a model with tunable entanglement entropy,} we replace the $\sin(k_y)$ and $\cos(k_y)$ functions in Eq.\ (\ref{eqn:extensionSimple}) by 
the {$C^{\infty}$ functions ${s_\mu},{c_\mu}:(-\pi,\pi]\rightarrow\RR$, }
\begin{eqnarray}
{s_\mu}(k_y) &=&\begin{cases}
 \text{sgn}(k_y)\text{e}^{-{k_y^{-2}}- {(\lvert k_y\rvert - \mu)^{-2}}+{8}/{\mu^2}} & \lvert k_y\rvert<\mu,\\
  0 & \mu<\lvert k_y\rvert<\pi,
  \end{cases}
  \end{eqnarray}
  \begin{eqnarray}
 {{c_\mu}}(k_y)& =&\begin{cases}
  ({1-{s_\mu}(k_y)^2})^{1/2} & \lvert k_y\rvert<\mu/2,\\
   - ({1-{s_\mu}(k_y)^2})^{1/2} & \lvert k_y\rvert>\mu/2,
 \end{cases}
\end{eqnarray}
where $\mu\in (0,\pi]$
is a real parameter and ${s_\mu},{{c_\mu}}$  are $2\pi$ periodically continued outside of the interval {$\left(-\pi,\pi\right]$}, to formally define a lattice model with unit lattice constant. These functions satisfy ${s_\mu}^2+{{c_\mu}}^2=1$ and have the same behavior under parity as $\sin$ and $\cos$, respectively. The substitution of $\sin(k_y)$ by ${s_\mu}(k_y)$ and $\cos(k_y)$ by ${{c_\mu}}(k_y)$ does not change the {instantaneous} 1D models at $k_y=0$ and $k_y=\pm \pi$. This modified model 
represents a smooth interpolation between the same 1D models and still has Chern number one. However, the Hamiltonian depends on $k_y$ only in the tunable interval ${(}0,\mu]$. Following our previous line of argumentation, the edge state contribution to the area law coefficient $\alpha$ hence becomes arbitrarily small for small $\mu$. However, the trivial 1D model $h(kx,\lvert ky\rvert >\mu)=\sin(k_x)\sigma_{{2}} +(2-\cos(k_x))\sigma_{{3}}$ still gives a non-topological contribution to the area law due to its dependence on $k_x$ that gives rise to delocalized states.  As a final step, we introduce a slightly modified dimensional extension, still with Chern number one, to the $k_x$-independent trivial 1D model  $\tilde h_\mu(kx,\lvert ky\rvert >\mu)=2 \sigma_{{3}}$ 
by defining
\begin{align}
&\tilde h_\mu(k_x,k_y)={s_\mu}(k_y) \sigma_{{1}} + \frac{1}{2}\left(1+{{c_\mu}}(k_y)\right)\sin(k_x)\sigma_{{2}}\nonumber\\
+&\bigl(1-{{c_\mu}}(k_y)-\frac{1}{2}\left(1+{{c_\mu}}(k_y)\right)\cos(k_x)\bigr)\sigma_{{3}}. 
\label{eqn:model}
\end{align}
The CI model (\ref{eqn:model}) is equal to the atomic insulator $h_0(k_x,k_y)=2\sigma_{{3}}$ for $\lvert k_y\rvert>\mu$, and is the key model for which we will demonstrate the tunability of the entanglement entropy. 

To further elucidate the properties of (\ref{eqn:model}), we compare the energy spectra on (long) cylinders as well as the Berry curvatures for a conventional Dirac model (\ref{eqn:extensionSimple}) and our family of CI Hamiltonians (\ref{eqn:model}) in Fig. \ref{fig:edgeberry}. For small values of $\mu$, the Berry curvature is strongly peaked around $k_y=\pm {\mu}/{2}$. The edge states exhibit a very steep slope around $k_y=\pm {\mu}/{2}$~and a plateau of width $\mu$~at zero energy between $k_y=-{\mu}/{2}$~and $k_y={\mu}/{2}$~which turns out to give the main contribution to the entanglement entropy. However, the whole range of low-energy states as a function of $k_y$~ decreases linearly with $\mu$~for small $\mu$~ and we hence expect the area law coefficient $\alpha$~of the entanglement entropy to do so as well, as confirmed by our numerical analysis (see Fig. \ref{fig:sscaling}) and complying with the analytical bounds to which we turn next.

\subsection{Upper and lower bounds to entanglement entropies} 
To assess that question analytically, we introduce a novel versatile tool to bound entanglement entropies allowing us to show that ground states of free fermionic systems for which correlations decay sufficiently rapidly  exhibit very little entanglement entropy. 
We first state the general result applied to a 1D system of length $L$, 
but it will be clear how to apply it to the above decoupled 2D situation. If $C(k_y)$ is the correlation matrix of a translationally invariant pure state
we say it decays with power $\beta>0$ whenever there exists a $c>0$ such that
\begin{equation}\label{decay1}
	|C(k_y)_{j,k}|\leq c \dist(j,k)^{-\beta},
\end{equation}
where $\dist$ is the distance in the lattice with periodic boundary conditions. 
For each $k_y$ {individually} one can then show the validity of an area law,
\begin{equation}
	\je{S_p(k_y) \leq c c_\beta},
\end{equation}	
where $c_\beta>0$ is a constant depending on $\beta$ only and $c$ is the constant of Eq.\ (\ref{decay1}): 
We find that  $\beta > 2$ is in fact sufficient to prove the validity of an area law in free fermionic models ({for details,
see the appendix}). 
The proof idea is to decompose the correlation matrix $C(k_y)$  for each $k_y$ into 
\begin{equation}
	C(k_y) = C^{(AB)}(k_y)+ M(k_y),
 \end{equation}	
where $C^{(AB)}(k_y)$ captures the uncorrelated situation between $A$ and its complement reflected by no entanglement at all. 
Then, one can use Weyl's perturbation theorem \cite{Bhatia2} to bound the {extent to which each of the eigenvalues of $ C(k_y)$
may be different from those of $M(k_y)$, a correlation matrix reflecting a product state. From a counting of the respective
eigenvalues and bounds to their magnitude, one arrives at a bound to \je{$S_p(k_y)$} from knowledge about the decay of correlations alone.}

{What remains to be seen is how one can derive the decay behavior of Eq.\ (\ref{decay1}) from the given dispersion relation of the model.
Here, an elegant tool comes into play, using a machinery of harmonic analysis: The decay of correlations is obtained from a suitable Fourier series
of the dispersion relation. One can derive such a decay, however, purely from knowledge about derivatives of the dispersion relation:}
$c$ is obtained from integrals over 
absolute values of third derivatives of 
dispersion relations. In this way, one arrives at the result with little computation, albeit in a fully rigorous way: It is clear from the 
dispersion relation of our model (\ref{eqn:model}) that these integrals over third derivatives can be made arbitrarily small (for details, see the 
appendix).
{Intuitively put, we hence make use of the freedom to appropriately tune the correlation decay in real space 
by altering the physical model to our desire, while keeping the topological features intact.}

\begin{observation}[Very low entanglement in Chern insulators] For any $\alpha>0$ \je{and any $p\geq 1$}, there are
two-band Chern insulator models on $L\times L$ tori, $L\geq L_0$, $L_0$ suitable, such that
the entanglement entropy of the bi-sected system satisfies
\begin{equation}
\je{S_p{(L)}\leq \alpha L} .
\end{equation}
\end{observation}
For this to be valid, one merely has to pick $\mu>0$ sufficiently small, as $\alpha$ will be monotone decreasing with $\mu$ and will approach zero.
Since in the partially decoupled situation one can lower bound the entanglement entropy of each 1D system by a continuous function 
$k_y\mapsto f(k_y)$ for the translationally invariant gapped models considered here, it is easy to see that in the thermodynamic limit $L\rightarrow \infty$ and with a concomitant 
refinement in momentum space, 
 the entanglement entropy has to grow linearly
in $L$, unless the ground state is obtained for a trivial, $\mathbf k$-independent, model with constant $\mathbf d$.
Such a state is however topologically trivial and separated from any Chern insulator by a closing of the bulk gap.

\begin{observation}[Non-trivial area laws in Chern insulators] 
For any two-band Chern insulator model on $L\times L$ tori \je{and any $p\geq 1$},
there exists an $\alpha>0$ and a $L_0$
such that the entanglement entropy of the bi-sected system satisfies 
\je{$S_p{(L)}\geq \alpha L$ for $L\geq L_0$}.
\end{observation}

\subsection{Numerical analysis}
{To complement the analytical considerations, we have performed an extensive numerical analysis which we briefly report here.} All of our numerical results are fully consistent with the rigorous results and also confirm that the actual entanglement boundary scales similarly with $\mu$ as the analytical bound does in the limit of large systems sizes. In particular, for small $\mu$ and large $L$ we find that $S_1(L)/L$ is indeed directly proportional to $\mu$ to very high precision, as is illustrated in Fig.\ \ref{fig:sscaling} for $L=60.000$ 
\cite{Remark}. 

\begin{figure}[t]
\includegraphics[width=1.0\linewidth]{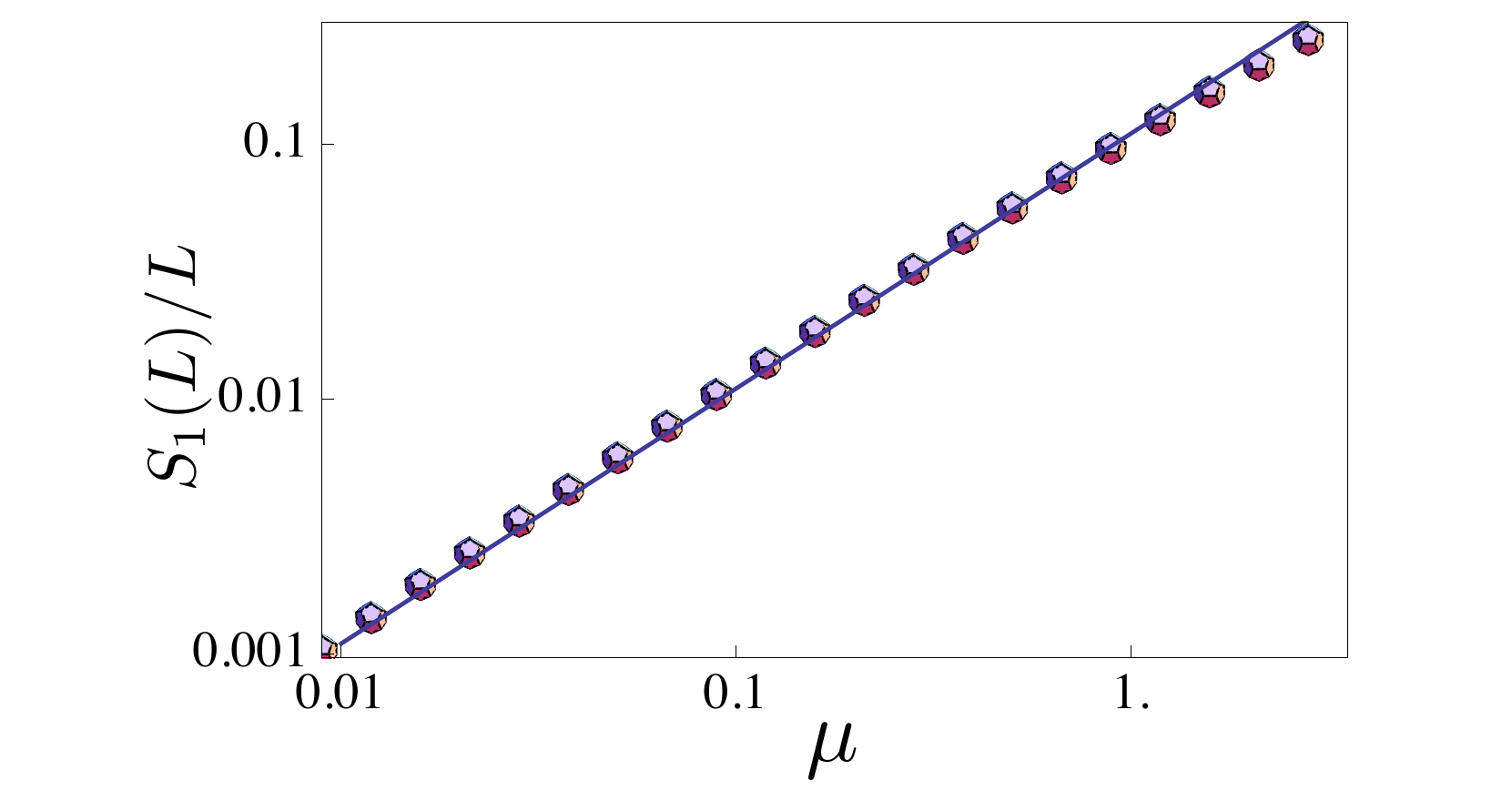}
\caption{(Color online) {Entanglement area law manifested in the large $L$ limit of \ejb{$S_1(L)/L$} as a function of $\mu$ on a log-log scale. For {sufficiently small} 
$\mu$, the numerically obtained values of \ejb{$S_1(L)/L$} are proportional to $\mu$ to an extraordinary precision, confirming our conclusion that $\lim_{L\rightarrow\infty} \ejb{S_1(L)/L} \rightarrow 0$ as $\mu\downarrow 0$. The blue line $0.11\mu$ is included as a guide to the eye.}}
\label{fig:sscaling}
\end{figure}

\section{Conclusions and discussion}
\label{sec:conclusion}
We have demonstrated the close correspondence between the area law entanglement scaling of topological insulators in two dimensions and their topologically protected edge states. In particular, we have shown that the {entanglement entropy over a cut}, 
while being topologically protected from assuming the value zero, is non-universal, and in fact arbitrarily tunable. While the analysis focused on a model with 
broken time reversal symmetry and Chern number $C=1$, our construction immediately generalizes to models with arbitrary 
Chern numbers and to time-reversal symmetric $Z_2$ topological insulators in 2D: 
{Models with arbitrary Chern numbers are obtained by $k_x\mapsto Nk_x$, with integer $N$, which directly leads to $C\mapsto N C$ and 
\ejb{$S_p(L)\mapsto N S_p(L)$}. A time-reversal symmetric, \ejb{$\mathbb Z_2$} topological insulator with tunable entanglement consists of two time-reversed copies of our model in
Eq.\ (\ref{eqn:model})}.
{We {also} find that our analysis can be {generalized} to higher
dimensions. 

For the particular case of symmetry protected topological states in 1D systems, there is a finite lower bound to the entanglement entropy.} 
Co-occurring with our work, Ref.\ \cite{Sondhi} independently 
concluded that the entanglement spectrum of Chern insulators is non-universal using very different means. In our rigorous framework, this is natural as 
each Renyi entropy has been explicitly proven to be non-universal, and taken together, the (integer) Renyi entropies uniquely determine the complete entanglement spectrum. It 
is common practice to infer topological properties for counting the number of low lying `entanglement energies' as a function of the transverse momentum and comparing it with predictions from conformal field theory. In our model the entanglement is `switched off' at an arbitrarily chosen transverse momentum, thus inferring the topology of the ground state in this fashion is impossible in any finite-size numerical investigation.

Our results have notable implications for numerics in the context of tensor networks approaches to Chern insulators \cite{GPEPS,Dubail,Beri}.
In particular, our finding that topological phases can have 
{very} low entanglement  
is encouraging for the simulation of interacting topological phases using entanglement based approaches. Although the weakly entangled topological insulators introduced here have a 
peaked Berry curvature, it is becoming increasingly clear that interesting strongly correlated phases can exist far beyond the idealized Landau level situation with a constant Berry curvature \cite{EmilReview}. Having a lattice model with a tunable Berry curvature while maintaining a sizable band gap is likely to bring new insights. {We hope our results stimulate such further work.}

\subsection*{Acknowledgements}
{We thank E.\ Ardonne, D.\ Haldane, and M.\ Hermanns 
for inspiring discussions  and R.\ Thomale 
and P.\ Zanardi for comments.} 
JCB was sponsored {by the} Swedish Research Council (VR)
and the ERC Synergy Grant UQUAM, {JE by
the EU (SIQS, RAQUEL, COST), the ERC (TAQ), the FQXi, and the BMBF (QuOReP),}
and EJB is supported by the Emmy Noether program (BE 5233/1-1) of the {DFG}.

\section*{Appendix}

In this appendix, we discuss the methods used in order to formulate the results presented in the main text. The material presented here is 
not needed in order to understand the conclusions of the main text. Since new techniques 
are introduced here, however, we present them in great detail. We also present a figure that provides further intuition to the argument.

\maketitle

\subsection{Novel upper bounds for entanglement entropies}

{We here introduce a novel method to upper bound entanglement entropies in translationally invariant 1D free 
fermionic systems of $L_x$ sites, with $L_x$ even for simplicity of notation. This argument stands in some tradition with statements linking correlations
to entanglement entropies in free bosonic systems \cite{Cramer} or in general spin models, where an exponential decay of correlations is required \cite{Brandao}.
Here, a mere slow algebraic decay is sufficient. }

{The basic idea is to decompose the correlation matrix into a part of a direct sum of parts
relating to $A$ and its complement -- from which the entanglement entropy can be computed -- and a remaining part that reflects the quickly decaying 
correlations. Naive bounds on this remaining part will not be sufficient. Weyl's perturbation theorem \cite{Bhatia2}, however, will be the instrument that allows
to capture the decay and the precise number of spectral values of this remaining part. This argument is expected to be of use also in other contexts where
block-Toeplitz-methods \cite{Its} are inapplicable or too tedious.
The $2L_x\times 2L_x$ circulant correlation matrix is denoted by $C$. 
Again, we say it decays with power $\beta>0$ whenever there exists an $c>1$ such that
\begin{equation}\label{decay}
	|C_{j,k}|\leq c \dist(j,k)^{-\beta},
\end{equation}
where $\dist$ is the distance in the lattice with periodic boundary conditions. Of course, this is true in particular if the correlations decay exponentially with the distance.
\je{For generality, we consider arbitrary Renyi entropies $S_p$ with $p\geq 1$, with $S_1=S$ being the standard von-Neumann entropy. Again, as is well known, all positive integer Renyi
entropies uniquely specify the entire entanglement spectrum.}
}

\begin{theorem}[Upper bounds to entanglement entropies] Consider the ground state of a free fermionic 
translationally invariant two-banded system of even length $L_x$, with a filled lower and empty upper band, 
with correlations decaying as in Eq.\ (\ref{decay}). 
If $\beta>2$, then each (Renyi) entanglement entropy satisfies the area law 
$S_p\leq c c_\beta$ for a suitable constant $c_\beta>0$.
\end{theorem}

{
{\it Proof.} 
{The latter constant $c_\beta$ does not depend on the system size.}
Denote with $ \xi_j^\downarrow$ the $j$-th eigenvalue of a matrix in non-increasing order. 
Denote with $C^{(AB)}$ the sub-matrix of $C$ that is obtained from $C$ by the pinching for which all 
correlations between $A$ and the complement $B$ vanish. Since the state is pure, the spectra of the submatrices associated with $A$ and its complement will be identical, 
and hence
the \je{(Renyi)}-entanglement entropy can \je{for any $p\geq 1$} be written as
\begin{equation}
	S_p = \sum_{j=1}^{2 L_x} h_p(\xi_j^\downarrow (C^{(AB)}))	,
\end{equation}
in terms of the \je{family of} entropy \je{functions} $h_p:[0,1]\rightarrow[0,1]$ defined as 
\begin{equation}
	\je{
	h_p(x)=\frac{1}{1-p}  (\log_2 (x^p+(1-x)^p ).
	}
\end{equation}
The first and key step will be a  consequence of a proper use of Weyl's perturbation theorem \cite{Bhatia2}. 
Since the ground state is unique and a pure state,
the spectral values are 
all contained in $\{1,0\}$,
that is, 
\begin{equation}	
	\xi_j^\downarrow( C )= 1 
\end{equation}	
for $j=1,\dots, L_x$ and  
\begin{equation}
 \xi_j^\downarrow( C )= 0
 \end{equation}
 for $j=L_x+1,\dots, 2 L_x$: This reflects the lower band being filled 
and the upper being empty.
The remaining part is referred to as $M$, so that 
\begin{equation}
	C= C^{(AB)}+M.
\end{equation}
$M$ reflects the decaying correlations in the ground state.
We now make use of Weyl's perturbation theorem: We find for the largest $L_x$ (twice degenerate) eigenvalues of
$C^{(AB)}$ 
\begin{equation}
	1- \xi_j^\downarrow(C^{(AB)})\leq \xi_j^\downarrow (M), \,j=1,\dots, L_x,
\end{equation}
and for the smallest eigenvalues 
\begin{equation}
	 \xi_j^\downarrow(C^{(AB)})\leq \xi_{j-L_x}^\downarrow (M),\,\,j=L_x+1,\dots, 2 L_x.
\end{equation}
That is to say, both the large eigenvalues close to $1$ and the small ones close to $0$ are only slightly perturbed by the same eigenvalues of $M$.
This implies that, using the monotonicity of $h$ on $[0,1/2]$,
\je{
\begin{eqnarray}
	S_p &=& \sum_{j=1}^{L_x} h_p(\xi^\downarrow_j(C^{(AB)})) + \sum_{j=L_x+1}^{2 L_x} h_p(\xi^\downarrow_j(C^{(AB)}))\nonumber\\
	&\leq& 2 \sum_{j=1}^{L_x} h_p(\xi_j^\downarrow (M)).
\end{eqnarray}
}
We now again employ Weyl's perturbation theorem, but now to $M$: We reveil hence the structure of eigenvalues of $M$, that they rapidly decay in the same way as the correlation matrix
entries decay. Acknowledging that for each of the $4$ sub-blocks of $M$ one encounters rapidly decaying correlations,
and using that $h_p(1/2)=1$ is the maximum value of the entropy function,
one finds
\je{
\begin{eqnarray}
	S_p &\leq& 	8  \sum_{j=1}^\infty h_p(\min(j c j^{-\beta},1/2)),
\end{eqnarray}
} again giving rise to an upper bound by extending the sum to $L_x$ by one to $\infty$. We now use that 
\je{
\begin{equation}
	h_p(cx)\leq \max(1,p) c h_p(x) 
\end{equation}
}
for all $c>1 $ and all $x\in[0,1]$ \je{such that $cx\in [0,1]$.}
\je{What is more, it is easy to see that 
\begin{equation}\label{Use}
	 \log(1+x)\leq x
\end{equation}
for $x>0$. This means that }	
\je{
\begin{eqnarray}
	S_p  &\leq& 	8  \max(1,p)c \sum_{j=1}^\infty  h_p(  j^{1-\beta})=: c c_{p,\beta},
\end{eqnarray}
for a suitable \je{$c_{p,\beta}>0$}, when $\beta>2$, as the infinite sum then converges for $p\geq 1$. This can be easily seen 
using Eq.\ (\ref{Use}), employing the fact that
\begin{equation}
	\sum_{j=1}^\infty j^{-(\beta-1)p}<\infty
\end{equation}
whenever $(\beta-1)p>1$.}
\proofend
}

Note that {the same argument is also applicable for any number of bands and is stated for a two-banded model merely for simplicity of notation. {The above proof
is generally still valid as is, the only modification being that the pre-factor will linearly grow with the number of bands considered.}}

\subsection{Harmonic analysis and correlations decay}

{The actual decay behavior of the correlations can here be determined using {tools of} harmonic analysis \cite{HarmonicAnalysis}.}

\begin{lemma}[Fourier components \cite{HarmonicAnalysis}]Let $\bar{f}:\RR\rightarrow\RR$ be a $2\pi$-periodic three-times differentiable function such that
$\bar{f}^{(3)}$ is absolutely continuous, then the Fourier coefficients will for all $j$ be bounded from above by
\begin{equation}\label{bound}
	|f_j |\leq \frac{c}{|j|^3} ,\, c:= \int_{-\pi}^{\pi} |\bar{f}^{(3)}(x)|dx.
\end{equation}
\end{lemma}
{The choice to express the bound in terms of third derivatives is done for convenience only; higher derivatives would have been applicable as well.}

\subsection{Application to thermodynamic limits of dispersion relations}

{These bounds can most conveniently be applied to the situation where one considers instead of an $L\times L$ lattice with toroidal boundary conditions
an $L_x\times L_y$ lattice with the same boundary conditions. It is easy to see that by first considering the limit $L_x\rightarrow\infty$ and then the limit $L_y\rightarrow \infty$,
one can obtain a rigorous bound on 
	$\lim_{L\rightarrow \infty} S(L)/L$ for the original $L\times L$-lattices at hand.
In this way, one can for each $k_y$ discuss an appropriate 
model in the thermodynamic, which simplifies the argument considerably. More precisely put, 
 this is a consequence of the fact that there exists an $\gamma>0$ such that for $L_x\times L_y$-lattices, 
	for each $k_y$ the entanglement entropy is shown to satisfy
	\begin{equation}
		S(k_y,L_x,L_y)\leq \gamma .
		\end{equation}
		Naturally, we can separate the limits of large $L_x$ and $L_y$ in order to
	simplify the discussion.}
{In the light of this discussion,
we define for each $k_y$ the functions
$\bar {\cal C}_{\uparrow,\uparrow}(k_y):(-\pi,\pi]\rightarrow \RR$ and $\bar {\cal C}_{\downarrow,\downarrow}(k_y):(-\pi,\pi]\rightarrow \RR$ as
the continuum limits of $\bar C_{\uparrow,\uparrow}$ and $\bar C_{\downarrow,\downarrow}$ for $L_x\times L_y$ lattices in the limit $L_x\rightarrow \infty$.
The real space correlation matrices are then obtained by invoking a Fourier transform, rendering the above Lemma on the Fourier coefficients of $2\pi$-periodic functions
applicable.
}

\subsection{Discussion of Chern insulator models considered}

{In this subsection we discuss the dispersion relations $\bar {\cal C}_{\uparrow,\uparrow}(k_y)$ for the above model at hand stated in the main text (compare also Fig.\ 1).
Equipped with the above
powerful tools, we will see that we can arrive at our conclusion almost without computation.
We find that 
\begin{eqnarray}
	\bar {\cal C}_{\uparrow,\uparrow}(x,k_y) = 0
\end{eqnarray}
for all $x\in(-\pi,\pi]$ and $\mu<k_y<\pi$. For $k_y\in[-\mu,\mu]$, we find that $x\mapsto \bar {\cal C}_{\uparrow,\uparrow}(x,k_y)$ is a $C^\infty$-function
with uniformly bounded third derivative. 
Similarly, one can argue about $\bar {\cal C}_{\downarrow,\downarrow}(x,k_y)$, since
\begin{eqnarray}
	\bar {\cal C}_{\downarrow,\downarrow}(x,k_y) = 0
\end{eqnarray}
for all $x\in(-\pi,\pi]$ and $\mu<k_y<\pi$. Again, for $k_y\in[-\mu,\mu]$,  the function $x\mapsto \bar {\cal C}_{\downarrow,\downarrow}(x,k_y)$ is a $C^\infty$-function
with uniformly bounded third derivative. The off-diagonal elements of the correlation matrix must decay at least as far as the main diagonal elements, as the correlation matrix is positive semi-definite.}

\subsection{Entanglement area laws}

{Equipped with the tools developed, Observation 1 follows immediately: 
Considering an $L_x\times L_y$ lattice,
in the limit $L_x\rightarrow \infty$, for values
$\mu<k_y<\pi$, there is no contribution to the entanglement entropy, while for $0<|k_y|<\mu$ the contribution is bounded from above by a constant: This is a consequence of Theorem 3 and 
Lemma 1. Using the above argument on the convergence for $L\times L$ lattices defined on the torus, we can conclude
that $\mu\rightarrow 0$, $S(L)/L$ converges to zero. 
This proves the validity of Observation 1. 
}

{In fact, an even
stronger statement follows, one that is also corroborated by the numerical analysis presented in the main text: The convergence to zero is essentially linear in $\mu$:
Precisely put, using the above machinery, it follows that there is a constant $c	>0$ such that
\begin{equation}
	\lim_{\mu\rightarrow 0} \frac{1}{\mu}\lim_{L\rightarrow\infty} \frac{S(L)}{L}\leq c.
\end{equation}
Intuitively speaking, this follows from the observation that along the $k_y$ direction, the number of contributing terms would shrink linearly in $\mu$, each of which is bounded from above
by a constant. With the tools developed, this is a conclusion that can be reached with little calculation.
}

\subsection{Lower bound}
{
We finally briefly discuss Observation 2, the lower bound to the entanglement entropy. It is clear that any continuous non-zero function $f:(-\pi,\pi]\rightarrow \RR^+$ with
\begin{equation}
	f(k_y)\leq S(k_y)
\end{equation}
will serve as a tool to show that Observation 2 is valid: Consider the partially decoupled situation along the $k_y$ direction.
Let $I\subset(-\pi,\pi]$ be an interval in the momentum along the cut with $f(k_y)>\epsilon$ for a suitable $\epsilon>0$, then in the thermodynamic limit $L_y\rightarrow\infty$,
one will encounter a contribution for the entanglement entropy bounded from below by $\alpha L$ for a suitable $\alpha$.
For the function $f$, several candidates are meaningful. 
For example, denote with $D^{(AB)}$ the correlation matrix of the two pairs of two sites each of the lattice immediately to the left or 
the right of the cut for the periodic boundary conditions chosen,
and let $D^{(A)}$ be the submatrix of $D^{(AB)}$ of only two sites belonging to $A$. Then
\begin{equation}\label{Mutual}
	S(k_y)\geq f(k_y):= \sum_j 2 h(\xi_j^\downarrow(D^{(A)}))-
	 \sum_j h(\xi_j^\downarrow(D^{(AB)})),
\end{equation}
(the mutual information),
which is always strictly positive unless the state is a product state, and the correlation matrix $D^{(AB)}$ is a continuous function of $k_y$
for the gapped models considered.
}


\begin{thebibliography}{99}
\expandafter\ifx\csname natexlab\endcsname\relax\def\natexlab#1{#1}\fi
\expandafter\ifx\csname bibnamefont\endcsname\relax
  \def\bibnamefont#1{#1}\fi
\expandafter\ifx\csname bibfnamefont\endcsname\relax
  \def\bibfnamefont#1{#1}\fi
\expandafter\ifx\csname citenamefont\endcsname\relax
  \def\citenamefont#1{#1}\fi
\expandafter\ifx\csname url\endcsname\relax
  \def\url#1{\texttt{#1}}\fi
\expandafter\ifx\csname urlprefix\endcsname\relax\def\urlprefix{URL }\fi
\providecommand{\bibinfo}[2]{#2}
\providecommand{\eprint}[2][]{\url{#2}}

\bibitem[{\citenamefont{Klitzing et~al.}(1980)\citenamefont{Klitzing, Dorda,
  and Pepper}}]{Klitzing1980}
\bibinfo{author}{\bibfnamefont{K.~v.} \bibnamefont{Klitzing}},
  \bibinfo{author}{\bibfnamefont{G.}~\bibnamefont{Dorda}}, \bibnamefont{and}
  \bibinfo{author}{\bibfnamefont{M.}~\bibnamefont{Pepper}},
  \bibinfo{journal}{Phys.\ Rev.\ Lett.} \textbf{\bibinfo{volume}{45}},
  \bibinfo{pages}{494} (\bibinfo{year}{1980}).

\bibitem[{\citenamefont{Stormer et~al.}(1983)\citenamefont{Stormer, Chang,
  Tsui, Hwang, Gossard, and Wiegmann}}]{StormerFQH}
\bibinfo{author}{\bibfnamefont{H.~L.} \bibnamefont{Stormer}},
  \bibinfo{author}{\bibfnamefont{A.}~\bibnamefont{Chang}},
  \bibinfo{author}{\bibfnamefont{D.~C.} \bibnamefont{Tsui}},
  \bibinfo{author}{\bibfnamefont{J.~C.~M.} \bibnamefont{Hwang}},
  \bibinfo{author}{\bibfnamefont{A.~C.} \bibnamefont{Gossard}},
  \bibnamefont{and} \bibinfo{author}{\bibfnamefont{W.}~\bibnamefont{Wiegmann}},
  \bibinfo{journal}{Phys.\ Rev.\ Lett.} \textbf{\bibinfo{volume}{50}},
  \bibinfo{pages}{1953} (\bibinfo{year}{1983}).

\bibitem[{\citenamefont{Laughlin}(1981)}]{Laughlin1981}
\bibinfo{author}{\bibfnamefont{R.~B.} \bibnamefont{Laughlin}},
  \bibinfo{journal}{Phys.\ Rev.\ B} \textbf{\bibinfo{volume}{23}},
  \bibinfo{pages}{5632} (\bibinfo{year}{1981}).

\bibitem[{\citenamefont{Thouless et~al.}(1982)\citenamefont{Thouless, Kohmoto,
  Nightingale, and den Nijs}}]{TKNN1982}
\bibinfo{author}{\bibfnamefont{D.~J.} \bibnamefont{Thouless}},
  \bibinfo{author}{\bibfnamefont{M.}~\bibnamefont{Kohmoto}},
  \bibinfo{author}{\bibfnamefont{M.~P.} \bibnamefont{Nightingale}},
  \bibnamefont{and} \bibinfo{author}{\bibfnamefont{M.}~\bibnamefont{den Nijs}},
  \bibinfo{journal}{Phys.\ Rev.\ Lett.} \textbf{\bibinfo{volume}{49}},
  \bibinfo{pages}{405} (\bibinfo{year}{1982}).

\bibitem[{\citenamefont{Laughlin}(1983)}]{LaughlinState}
\bibinfo{author}{\bibfnamefont{R.~B.} \bibnamefont{Laughlin}},
  \bibinfo{journal}{Phys.\ Rev.\ Lett.} \textbf{\bibinfo{volume}{50}},
  \bibinfo{pages}{1395} (\bibinfo{year}{1983}).

\bibitem[{\citenamefont{Wen}(1990)}]{WenTO}
\bibinfo{author}{\bibfnamefont{X.-G.} \bibnamefont{Wen}},
  \bibinfo{journal}{Int. J. Mod. Phys. B} \textbf{\bibinfo{volume}{4}},
  \bibinfo{pages}{239} (\bibinfo{year}{1990}).

\bibitem[{\citenamefont{Anderson}(1997)}]{AndersonBasic}
\bibinfo{author}{\bibfnamefont{P.~W.} \bibnamefont{Anderson}},
  \emph{\bibinfo{title}{{Basic notions of condensed matter physics}}}
  (\bibinfo{publisher}{Perseus}, \bibinfo{year}{1997}).

\bibitem[{\citenamefont{Eisert et~al.}(2010)\citenamefont{Eisert, Cramer, and
  Plenio}}]{JensReview}
\bibinfo{author}{\bibfnamefont{J.}~\bibnamefont{Eisert}},
  \bibinfo{author}{\bibfnamefont{M.}~\bibnamefont{Cramer}}, \bibnamefont{and}
  \bibinfo{author}{\bibfnamefont{M.~B.} \bibnamefont{Plenio}},
  \bibinfo{journal}{Rev. Mod. Phys.} \textbf{\bibinfo{volume}{82}},
  \bibinfo{pages}{277} (\bibinfo{year}{2010}).

\bibitem[{\citenamefont{{Schollw{\"o}ck}}(2011)}]{DMRGMPS}
\bibinfo{author}{\bibfnamefont{U.}~\bibnamefont{{Schollw{\"o}ck}}},
  \bibinfo{journal}{Ann. Phys.} \textbf{\bibinfo{volume}{326}},
  \bibinfo{pages}{96} (\bibinfo{year}{2011}).

\bibitem[{\citenamefont{Kitaev and Preskill}(2006)}]{KitaevPreskill}
\bibinfo{author}{\bibfnamefont{A.}~\bibnamefont{Kitaev}} \bibnamefont{and}
  \bibinfo{author}{\bibfnamefont{J.}~\bibnamefont{Preskill}},
  \bibinfo{journal}{Phys.\ Rev.\ Lett.} \textbf{\bibinfo{volume}{96}},
  \bibinfo{pages}{110404} (\bibinfo{year}{2006}).

\bibitem[{\citenamefont{Levin and Wen}(2006)}]{LevinWenEntropy}
\bibinfo{author}{\bibfnamefont{M.}~\bibnamefont{Levin}} \bibnamefont{and}
  \bibinfo{author}{\bibfnamefont{X.-G.} \bibnamefont{Wen}},
  \bibinfo{journal}{Phys.\ Rev.\ Lett.} \textbf{\bibinfo{volume}{96}},
  \bibinfo{pages}{110405} (\bibinfo{year}{2006}).

\bibitem{Zanardi}
	A.\ Hamma, R.\ Ionicioiu, and P.\ Zanardi, Phys.\ Lett.\ A {\bf 22}, 337 (2005).

\bibitem{Zanardi2}
	A.\ Hamma, R.\ Ionicioiu, and P.\ Zanardi, Phys.\ Rev.\ A {\bf 71}, 022315 (2005).

\bibitem[{\citenamefont{Haldane}(1988)}]{QAH}
\bibinfo{author}{\bibfnamefont{F.~D.~M.} \bibnamefont{Haldane}},
  \bibinfo{journal}{Phys.\ Rev.\ Lett.} \textbf{\bibinfo{volume}{61}},
  \bibinfo{pages}{2015} (\bibinfo{year}{1988}).

\bibitem[{\citenamefont{Chen et~al.}(2010)\citenamefont{Chen, Gu, and
  Wen}}]{WenLU}
\bibinfo{author}{\bibfnamefont{X.}~\bibnamefont{Chen}},
  \bibinfo{author}{\bibfnamefont{Z.-C.} \bibnamefont{Gu}}, \bibnamefont{and}
  \bibinfo{author}{\bibfnamefont{X.-G.} \bibnamefont{Wen}},
  \bibinfo{journal}{Phys.\ Rev.\ B} \textbf{\bibinfo{volume}{82}},
  \bibinfo{pages}{155138} (\bibinfo{year}{2010}).

\bibitem[{\citenamefont{Kane and Mele}(2005{\natexlab{a}})}]{KaneMele2005a}
\bibinfo{author}{\bibfnamefont{C.~L.} \bibnamefont{Kane}} \bibnamefont{and}
  \bibinfo{author}{\bibfnamefont{E.~J.} \bibnamefont{Mele}},
  \bibinfo{journal}{Phys.\ Rev.\ Lett.} \textbf{\bibinfo{volume}{95}},
  \bibinfo{pages}{226801} (\bibinfo{year}{2005}{\natexlab{a}}).

\bibitem[{\citenamefont{Kane and Mele}(2005{\natexlab{b}})}]{KaneMele2005b}
\bibinfo{author}{\bibfnamefont{C.~L.} \bibnamefont{Kane}} \bibnamefont{and}
  \bibinfo{author}{\bibfnamefont{E.~J.} \bibnamefont{Mele}},
  \bibinfo{journal}{Phys.\ Rev.\ Lett.} \textbf{\bibinfo{volume}{95}},
  \bibinfo{pages}{146802} (\bibinfo{year}{2005}{\natexlab{b}}).

\bibitem[{\citenamefont{Bernevig et~al.}(2006)\citenamefont{Bernevig, Hughes,
  and Zhang}}]{BHZ2006}
\bibinfo{author}{\bibfnamefont{B.~A.} \bibnamefont{Bernevig}},
  \bibinfo{author}{\bibfnamefont{T.~L.} \bibnamefont{Hughes}},
  \bibnamefont{and} \bibinfo{author}{\bibfnamefont{S.-C.} \bibnamefont{Zhang}},
  \bibinfo{journal}{Science} \textbf{\bibinfo{volume}{314}},
  \bibinfo{pages}{1757} (\bibinfo{year}{2006}).

\bibitem[{\citenamefont{Schnyder et~al.}(2008)\citenamefont{Schnyder, Ryu,
  Furusaki, and Ludwig}}]{Schnyder2008}
\bibinfo{author}{\bibfnamefont{A.~P.} \bibnamefont{Schnyder}},
  \bibinfo{author}{\bibfnamefont{S.}~\bibnamefont{Ryu}},
  \bibinfo{author}{\bibfnamefont{A.}~\bibnamefont{Furusaki}}, \bibnamefont{and}
  \bibinfo{author}{\bibfnamefont{A.~W.~W.} \bibnamefont{Ludwig}},
  \bibinfo{journal}{Phys.\ Rev.\ B} \textbf{\bibinfo{volume}{78}},
  \bibinfo{pages}{195125} (\bibinfo{year}{2008}).

\bibitem[{\citenamefont{Kitaev}(2009)}]{KitaevPeriodic}
\bibinfo{author}{\bibfnamefont{A.}~\bibnamefont{Kitaev}}, \bibinfo{journal}{AIP
  Conf. Proceedings} \textbf{\bibinfo{volume}{1134}}, \bibinfo{pages}{22}
  (\bibinfo{year}{2009}).

\bibitem[{\citenamefont{{Ryu} et~al.}(2010)\citenamefont{{Ryu}, {Schnyder},
  {Furusaki}, and {Ludwig}}}]{RyuLudwig}
\bibinfo{author}{\bibfnamefont{S.}~\bibnamefont{{Ryu}}},
  \bibinfo{author}{\bibfnamefont{A.~P.} \bibnamefont{{Schnyder}}},
  \bibinfo{author}{\bibfnamefont{A.}~\bibnamefont{{Furusaki}}},
  \bibnamefont{and} \bibinfo{author}{\bibfnamefont{A.~W.~W.}
  \bibnamefont{{Ludwig}}}, \bibinfo{journal}{New J. Phys.}
  \textbf{\bibinfo{volume}{12}}, \bibinfo{pages}{065010}
  (\bibinfo{year}{2010}).
  
  
\bibitem[{\citenamefont{Witten}(1983)}]{WZW}
\bibinfo{author}{\bibfnamefont{E.}~\bibnamefont{Witten}},
  \bibinfo{journal}{Nucl.\ Phys.\ B} \textbf{\bibinfo{volume}{223}},
  \bibinfo{pages}{422 } (\bibinfo{year}{1983}).

\bibitem[{\citenamefont{Qi et~al.}(2008)\citenamefont{Qi, Hughes, and
  Zhang}}]{QiTFT}
\bibinfo{author}{\bibfnamefont{X.-L.} \bibnamefont{Qi}},
  \bibinfo{author}{\bibfnamefont{T.~L.} \bibnamefont{Hughes}},
  \bibnamefont{and} \bibinfo{author}{\bibfnamefont{S.-C.} \bibnamefont{Zhang}},
  \bibinfo{journal}{Phys.\ Rev.\ B} \textbf{\bibinfo{volume}{78}},
  \bibinfo{pages}{195424} (\bibinfo{year}{2008}).

\bibitem{LiHaldane}  { H.~Li and F.~D.~M.~Haldane, Phys.\ Rev.\ Lett.\
 {\bf 101}, 010504 (2008).}

\bibitem{Qi}
{ X.-L. Qi, H. Katsura, and A. W. W. Ludwig, Phys. Rev.
Lett. {\bf 108}, 196402 (2012).}



\bibitem[{\citenamefont{Kapit and Mueller}(2010)}]{kapitmueller}
\bibinfo{author}{\bibfnamefont{E.}~\bibnamefont{Kapit}} \bibnamefont{and}
  \bibinfo{author}{\bibfnamefont{E.}~\bibnamefont{Mueller}},
  \bibinfo{journal}{Phys.\ Rev.\ Lett.} \textbf{\bibinfo{volume}{105}},
  \bibinfo{pages}{215303} (\bibinfo{year}{2010}).

\bibitem[{\citenamefont{Tang et~al.}(2011)\citenamefont{Tang, Mei, and
  Wen}}]{wenfci}
\bibinfo{author}{\bibfnamefont{E.}~\bibnamefont{Tang}},
  \bibinfo{author}{\bibfnamefont{J.-W.} \bibnamefont{Mei}}, \bibnamefont{and}
  \bibinfo{author}{\bibfnamefont{X.-G.} \bibnamefont{Wen}},
  \bibinfo{journal}{Phys.\ Rev.\ Lett.} \textbf{\bibinfo{volume}{106}},
  \bibinfo{pages}{236802} (\bibinfo{year}{2011}).

\bibitem[{\citenamefont{Sun et~al.}(2011)\citenamefont{Sun, Gu, Katsura, and
  Das~Sarma}}]{sunfci}
\bibinfo{author}{\bibfnamefont{K.}~\bibnamefont{Sun}},
  \bibinfo{author}{\bibfnamefont{Z.}~\bibnamefont{Gu}},
  \bibinfo{author}{\bibfnamefont{H.}~\bibnamefont{Katsura}}, \bibnamefont{and}
  \bibinfo{author}{\bibfnamefont{S.}~\bibnamefont{Das~Sarma}},
  \bibinfo{journal}{Phys.\ Rev.\ Lett.} \textbf{\bibinfo{volume}{106}},
  \bibinfo{pages}{236803} (\bibinfo{year}{2011}).

\bibitem[{\citenamefont{Neupert et~al.}(2011)\citenamefont{Neupert, Santos,
  Chamon, and Mudry}}]{neupertfci}
\bibinfo{author}{\bibfnamefont{T.}~\bibnamefont{Neupert}},
  \bibinfo{author}{\bibfnamefont{L.}~\bibnamefont{Santos}},
  \bibinfo{author}{\bibfnamefont{C.}~\bibnamefont{Chamon}}, \bibnamefont{and}
  \bibinfo{author}{\bibfnamefont{C.}~\bibnamefont{Mudry}},
  \bibinfo{journal}{Phys.\ Rev.\ Lett.} \textbf{\bibinfo{volume}{106}},
  \bibinfo{pages}{236804} (\bibinfo{year}{2011}).

\bibitem[{\citenamefont{{Bergholtz} and {Liu}}(2013)}]{EmilReview}
\bibinfo{author}{\bibfnamefont{E.~J.} \bibnamefont{{Bergholtz}}}
  \bibnamefont{and} \bibinfo{author}{\bibfnamefont{Z.}~\bibnamefont{{Liu}}},
  \bibinfo{journal}{Int. J. Mod. Phys. B} \textbf{\bibinfo{volume}{27}}, \bibinfo{pages}{1330017}  (\bibinfo{year}{2013}).

\bibitem[{\citenamefont{{Peschel}}(2003)}]{Peschel2003}
\bibinfo{author}{\bibfnamefont{I.}~\bibnamefont{{Peschel}}},
  \bibinfo{journal}{J. Phys. A}
  \textbf{\bibinfo{volume}{36}}, \bibinfo{pages}{L205} (\bibinfo{year}{2003}).
  
  \bibitem[{\citenamefont{Eisert}(2007)}]{JensStat}
\bibinfo{author}{\bibfnamefont{M.}~\bibnamefont{Cramer}},
\bibinfo{author}{\bibfnamefont{J.}~\bibnamefont{Eisert}} \bibnamefont{and}
  \bibinfo{author}{\bibfnamefont{M. B.}~\bibnamefont{Plenio}},
  \bibinfo{journal}{Phys.\ Rev.\ Lett.} \textbf{\bibinfo{volume}{98}},
  \bibinfo{pages}{220603} (\bibinfo{year}{2007}); M.\ Cramer, PhD thesis (Potsdam, Jan.\ 2007).
  
  \bibitem[{\citenamefont{{Eisert}}(2013)}]{JensAnalytical}
\bibinfo{author}{\bibfnamefont{H.}~\bibnamefont{Bernigau}},
\bibinfo{author}{\bibfnamefont{M. J.}~\bibnamefont{Kastoryano}} \bibnamefont{and}
  \bibinfo{author}{\bibfnamefont{J.}~\bibnamefont{Eisert}},
  \bibinfo{journal}{arXiv:1301.5646}.

\bibitem[{\citenamefont{Turner et~al.}(2010)\citenamefont{Turner, Zhang, and
  Vishwanath}}]{Turner2009}
\bibinfo{author}{\bibfnamefont{A.~M.} \bibnamefont{Turner}},
  \bibinfo{author}{\bibfnamefont{Y.}~\bibnamefont{Zhang}}, \bibnamefont{and}
  \bibinfo{author}{\bibfnamefont{A.}~\bibnamefont{Vishwanath}},
  \bibinfo{journal}{Phys.\ Rev.\ B} \textbf{\bibinfo{volume}{82}},
  \bibinfo{pages}{241102} (\bibinfo{year}{2010}).



\bibitem{Fidkowski}{L.\ Fidkowski, Phys.\ Rev.\ Lett.\ {\bf 104}, 130502 (2010).}

\bibitem{Turner2012}{A.\ M.\ Turner, Y.\ Zhang, R.\ S.\ K.\ Mong, and A.\ Vishwanath, 
Phys.\ Rev.\ B {\bf 85}, 165120 (2012).}

\bibitem{Hughes}{T.\ L.\ Hughes, E.\ Prodan, and B. A. Bernevig, Phys. Rev. B {\bf 83}, 
245132 (2011).}

\bibitem{FlatBandFootnote}
 	{This can be achieved by} {replacing} ${\mathbf d}({\mathbf k})\mapsto{\mathbf d}({\mathbf k})/(2\|  {\mathbf d}({\mathbf k}) \| )$.

\bibitem[{\citenamefont{Alexandradinata
  et~al.}(2011)\citenamefont{Alexandradinata, Hughes, and
  Bernevig}}]{Alexandradinata}
\bibinfo{author}{\bibfnamefont{A.}~\bibnamefont{Alexandradinata}},
  \bibinfo{author}{\bibfnamefont{T.~L.} \bibnamefont{Hughes}},
  \bibnamefont{and} \bibinfo{author}{\bibfnamefont{B.~A.}
  \bibnamefont{Bernevig}}, \bibinfo{journal}{Phys.\ Rev.\ B}
  \textbf{\bibinfo{volume}{84}}, \bibinfo{pages}{195103}
  (\bibinfo{year}{2011}).

\bibitem[{\citenamefont{Su et~al.}(1979)\citenamefont{Su, Schrieffer, and
  Heeger}}]{SSH}
\bibinfo{author}{\bibfnamefont{W.~P.} \bibnamefont{Su}},
  \bibinfo{author}{\bibfnamefont{J.~R.} \bibnamefont{Schrieffer}},
  \bibnamefont{and} \bibinfo{author}{\bibfnamefont{A.~J.}
  \bibnamefont{Heeger}}, \bibinfo{journal}{Phys.\ Rev.\ Lett.}
  \textbf{\bibinfo{volume}{42}}, \bibinfo{pages}{1698} (\bibinfo{year}{1979}).

\bibitem[{\citenamefont{Heeger et~al.}(1988)\citenamefont{Heeger, Kivelson,
  Schrieffer, and Su}}]{SSHReview}
\bibinfo{author}{\bibfnamefont{A.~J.} \bibnamefont{Heeger}},
  \bibinfo{author}{\bibfnamefont{S.}~\bibnamefont{Kivelson}},
  \bibinfo{author}{\bibfnamefont{J.~R.} \bibnamefont{Schrieffer}},
  \bibnamefont{and} \bibinfo{author}{\bibfnamefont{W.~P.} \bibnamefont{Su}},
  \bibinfo{journal}{Rev. Mod. Phys.} \textbf{\bibinfo{volume}{60}},
  \bibinfo{pages}{781} (\bibinfo{year}{1988}).

\bibitem[{\citenamefont{Zak}(1989)}]{ZakPol}
\bibinfo{author}{\bibfnamefont{J.}~\bibnamefont{Zak}}, \bibinfo{journal}{Phys.
  Rev. Lett.} \textbf{\bibinfo{volume}{62}}, \bibinfo{pages}{2747}
  (\bibinfo{year}{1989}).

\bibitem[{\citenamefont{Hatsugai}(2006)}]{HatsugaiQuantizedBerry}
\bibinfo{author}{\bibfnamefont{Y.}~\bibnamefont{Hatsugai}},
  \bibinfo{journal}{J.\ Phys.\ Soc.\ Japan}
  \textbf{\bibinfo{volume}{75}}, \bibinfo{pages}{123601}
  (\bibinfo{year}{2006}).


\bibitem{Sondhi}
 {A. Chandran, V. Khemani, and S. L. Sondhi, arXiv:1311.2946}.
  
\bibitem{GPEPS}
	{T.\ B.\ Wahl, H.-H.\ Tu, N.\ Schuch, and J.\ I.\ Cirac, Phys.\ Rev.\ Lett.\ {\bf 111}, 236805 (2013).}
	
\bibitem{Dubail}
	{J. Dubail and N. Read, arXiv:1307.7726.}

\bibitem{Beri}
	{B. Beri and N. R. Cooper, Phys.\ Rev.\ Lett. {\bf 106}, 156401 (2011).}
 
\bibitem{Remark}
	{In practice, we truncate the lattice in the direction perpendicular to the cut at $30$ sites for which the entropy is already converged essentially to machine precision.}


 \bibitem{Bhatia2}
    {R.\ Bhatia, {\it Matrix analysis} (Springer, Berlin, 1997).}


\bibitem{Cramer}
	{M.\ Cramer and J.\ Eisert, New J.\ Phys.\ {\bf 8}, 71 (2006).}
	
\bibitem{Brandao}
	{F.\ G.\ S.\ L.\ Brandao and M.\ Horodecki, Nature Phys.\   {\bf 9}, 721 (2013).}

\bibitem{Its}
	A.\ R.\ Its, B.-Q. Jin, and V.\ E.\ Korepin,
	quant-ph/0606178.


\bibitem{HarmonicAnalysis}
	{Y.\ Katznelson, {\it An introduction to harmonic analysis} (Cambridge Mathematical Library, 2004).}
	
		
  
\end{thebibliography}
\end{document}